\documentclass[aps,prl, showpacs,superscriptaddress]{revtex4}
\usepackage{dcolumn}
\usepackage{bm}
\usepackage[dvips]{graphics}
\begin{document}
\title{CP Measurement in Quantum Teleportation of Neutral Mesons}
\author{Yu Shi}
\affiliation{Department of Physics, Fudan University, Shanghai
200433, China}
\author{Yue-Liang Wu}
\affiliation{Institute of Theoretical Physics, Chinese Academy of
Sciences, Beijing 100080, China}

\begin{abstract}



\vspace{1cm}

{\bf Abstract}

\vspace{0.5cm}

Quantum teleportation using neutral pseudoscalar mesons shows novel
connections between particle physics and quantum information. The
projection basis, which is crucial in the teleportation process, is
determined by the conservation laws of particle physics, and is
different from the Bell basis, as in the usual case. Here we show
that one can verify the teleportation process by CP measurement.
This method significantly simplifies the high energy quantum
teleportation protocol. Especially, it is rigorous, and is
independent of whether CP is violated in weak decays. This method
can also be applied to general verification of
Einstein-Podolsky-Rosen correlations in particle physics.

\end{abstract}

\pacs{14.40.Aq, 13.90.+i, 03.67.Mn. }

\maketitle

\section*{1 Introduction}

It has been suggested that a  $K^0\bar{K}^0$ pair in an entangled
state can be produced from the strong decay of a $\phi$ meson
generated in electron-positron annihilation, or from a
proton-antiproton collision,  and that a similarly entangled
$B^0\bar{B}^0$ pair can be produced from $\Upsilon (4S)$
resonance~\cite{rev1,rev1a,rev2,rev2a}. Experimental results
consistent with the existence of entanglement has been demonstrated
in $K^0\bar{K}^0$ pairs produced in proton-antiproton annihilation
in the CPLEAR detector in CERN~\cite{cern}, in $K^0\bar{K}^0$ pairs
produced in $\phi$ decay in the KLOE detector in
DA$\Phi$NE~\cite{kloe,kloea}, as well as in $B^0\bar{B}^0$ pairs
produced in the Belle detector in the KEKB electron-positron
collider~\cite{go,goa}. Recently, the proposal was made to use
neutral kaons to realize such quantum information processes as
quantum teleportation and entanglement swapping in the regime of
high energy physics~\cite{shi}.  In this proposal,   the process is
verified by measuring the strangeness ratio of the particle to which
the state is teleported, or the strangeness asymmetry of the two
particles to which the entangled state is swapped. In this Letter,
we point out that one can also use measurement in the CP basis to
verify quantum teleportation or entanglement swapping, as well  as
to verify Einstein-Podolsky-Rosen correlations in general. The
proposal and the verification methods can also be implemented in
other neutral pseudoscalar mesons.

\section*{2 Entangled mesons}

For any neutral pseudoscalar meson $M^0$  and its antiparticle
$\bar{M}^0$, both with $J^P=0^-$, an entangled state can be produced
from a source with $J^{PC}=1^{--}$. $M^0$ can be $K^0=(d\bar{s})$,
$B^0=(d\bar{b})$, $\bar{B}_s^0=(b\bar{s})$ or
$\bar{D}^0=(u\bar{c})$. $|M^0\rangle$ and $|\bar{M}^0\rangle$ are
eigenstates of parity $P$ with eigenvalues $-1$ and also eigenstates
of its characteristic flavor  ${\cal F}$ with eigenvalues $1$ and
$-1$, respectively. ${\cal F}$ is strangeness for kaons, beauty for
$B$-mesons and charm for $D$-mesons, respectively. For $B_s$-mesons,
${\cal F}$ can be chosen to be strangeness or beauty (with minus
sign). We have $C|M^0\rangle = -|\bar{M}^0\rangle$ and
$C|\bar{M}^0\rangle = -|M^0\rangle$. Thus the eigenstates of $CP$
are
$$\begin{array}{rcl} |M_+\rangle & = &
\frac{1}{\sqrt{2}}(|M^0\rangle + |\bar{M}^0\rangle),\\
|M_{-}\rangle & = & \frac{1}{\sqrt{2}}(|M^0\rangle
-|\bar{M}^0\rangle),\end{array} $$ with eigenvalues $+1$ and $-1$,
respectively. Under weak interactions, the lifetime-mass eigenstates
are $|M_{SL}\rangle$ and $|M_{LH}\rangle$, with eigenvalues
$\lambda_{SL}=m_{SL}-i\Gamma_{SL}/2$ and
$\lambda_{LH}=m_{LH}-i\Gamma_{LH}/2$, where the subscript $SL$ means
short lifetime and light mass, while $LH$ means long lifetime and
heavy mass. Usually, the first subscript is used for kaons while the
second subscript is used for $B$-mesons, as kaons differ mainly in
lifetimes while $B$-mesons differ mainly in masses. Here we write
both subscripts for a general discussion. In terms of the proper
time $\tau$, the weak decay is described as $|M_{SL}(\tau)\rangle =
e^{-i\lambda_{SL} \tau} |M_{SL}\rangle$ and $|M_{LH}(\tau)\rangle =
e^{-i\lambda_{LH} \tau} |M_{LH}\rangle$. $|M_{SL}\rangle$ and
$|M_{LH}\rangle$ are related to $CP$ and strangeness eigenstates as
\begin{equation}
\begin{array}{lcl}
|M_{SL}\rangle & = & \frac{1}{\sqrt{1+|\epsilon|^2}}(|M_+\rangle
+\epsilon|M_-\rangle) \\ & = &
\frac{1}{\sqrt{|p|^2+|q|^2}}(p|M^0\rangle
+q|\bar{M}^0\rangle), \\
|M_{LH}\rangle & = & \frac{1}{\sqrt{1+|\epsilon|^2}}(|M_-\rangle
+\epsilon|M_+\rangle)\\ & = &
\frac{1}{\sqrt{|p|^2+|q|^2}}(p|M^0\rangle - q|\bar{M}^0\rangle),
\end{array} \label{ksl} \end{equation}
where $\epsilon$ is the parameter characterizing $CP$ violation,
$p=1+\epsilon$ and $q=1-\epsilon$. The magnitude of $\epsilon$ is of
the order of $10^{-3}$.

The entangled state with $J^{PC}=1^{--}$, written in terms of the
three bases respectively, is
\begin{eqnarray}
|\Psi_-\rangle & = & \frac{1}{\sqrt{2}}(|M^0\rangle|\bar{M}^0\rangle
-|\bar{M}^0\rangle|M^0\rangle)\\
&=&\frac{1}{\sqrt{2}}(|M_-\rangle|M_+\rangle-
|M_+\rangle|M_-\rangle) \label{cp} \\
&= & \frac{r}{\sqrt{2}}(|M_{LH}\rangle|M_{SL}\rangle-
|M_{SL}\rangle|M_{LH}\rangle). \label{r}
\end{eqnarray}
where $r=(|p|^2+|q|^2)/2pq = (1+|\epsilon|^2)/(1-\epsilon^2)$.

It is remarkable that in the CP basis, $|\Psi_-\rangle$ is also a
singlet strictly, and there is no dependence on the CP violation
parameter, which only appears in the expression in the lifetime-mass
basis. It should be noted that $|M_+\rangle$ and $|M_-\rangle$ are
orthogonal to each other, and are exactly distinguished by different
values of $CP$, while $|M_{SL}\rangle$ and $|M_{LH}\rangle$ are not
orthogonal to each other as a consequence of CP violation. This
makes it advantageous to use the CP basis, rather than the weak
basis, to do measurements.

In actual experiments, identification of $|M_+\rangle$ and
$|M_-\rangle$ is made by the modes of nonleptonic decays, i.e. a
$M_+$ decays to $2\pi$ while  a $M_-$ decays to $3\pi$. On the other
hand, although the physical states in free propagation are
$|M_{SL}\rangle$ and $|M_{LH}\rangle$, they cannot be exactly
identified due to CP violation. In the so-called passive measurement
of lifetime,  one identifies the mesons that decay between $\tau$
and $\tau+\Delta\tau$ as  $|M_{SL}\rangle$ and those decaying later
as $|M_{LH}\rangle$, by choosing the appropriate $\tau$ and
$\Delta\tau$~\cite{rev1,rev1a}. There is always nonzero
misidentification probability. Also, it relies on the significant
difference between $\Gamma_{SL}$ and $\Gamma_{LH}$ and  hence does
not apply to other cases such as $B$-mesons. Furthermore, one has to
exclude the semileptonic decays, which are in the flavor basis. To
do this, one has to look into the details of the decays. But the
nonleptonic decays are just in the CP basis. Therefore we would
rather abandon the identification in the lifetime-mass basis.

\section*{3 High energy quantum teleportation}

In the following we first review the proposal of high energy
teleportation and entanglement swapping. For the general case of
$M$-mesons, one can extend the calculation results in
Ref.~\cite{shi} by replacing $\Gamma_{S}$ and $\Gamma_{L}$ with
$i\lambda_{SL}$ and $i\lambda_{LH}$ respectively, as the mass
difference for kaons has been neglected there.

Suppose two mesons labeled by $a$ and $b$ are produced in state
$|\Psi_-\rangle$ at time $t=0$, in the laboratory frame that
coincides with the center of mass frame. At time $t$, the state
becomes
 \begin{equation}
|\Psi_{ab}(t)\rangle = M(t) |\Psi_-\rangle_{ab}, \label{abt}
\end{equation}
where $M(t) = \exp [-i(\lambda_{SL}+\lambda_{LH})\gamma^{-1}_{b}t]$.
$\gamma_i$ is the Lorentz factor $1/\sqrt{1-v_i^2}$ for particle $i$
with velocity $v_i$. It has been assumed that $\gamma_a=\gamma_b$. A
third kaon $c$ is generated at time $t_z$, with $|\Psi_c(t_z)\rangle
= \alpha |M^0\rangle_c +\beta |\bar{M}^0\rangle_c$;  thus
\begin{eqnarray}
|\Psi_c(t)\rangle & = & C(t) |M_{SL}\rangle_c + B(t) |M_{LH}\rangle_c \label{ctb} \\
&=& D(t) |M_+\rangle_c +E(t) |M_-\rangle_c, \label{cp}\\
& =& F(t) |M^0\rangle_c +G(t) |\bar{M}^0\rangle_c  \label{ct}
\end{eqnarray}
where \begin{widetext} $$C(t) =
\frac{\sqrt{1+|\epsilon|^2}}{\sqrt{2}}(\frac{\alpha}{1+\epsilon}
+\frac{\beta}{1-\epsilon}) e^{-i\lambda_{SL}\gamma^{-1}_c(t-t_z)},$$
$$B(t) =
\frac{\sqrt{1+|\epsilon|^2}}{\sqrt{2}}(\frac{\alpha}{1+\epsilon}
-\frac{\beta}{1-\epsilon}) e^{-i\lambda_{LH}\gamma^{-1}_c(t-t_z)},$$
$$D(t)= \frac{1}{\sqrt{2}}(\frac{\alpha}{1+\epsilon}
+\frac{\beta}{1-\epsilon}) e^{-i\lambda_{SL}\gamma^{-1}_c(t-t_z)}+
\frac{\epsilon}{\sqrt{2}}(\frac{\alpha}{1+\epsilon}
-\frac{\beta}{1-\epsilon}) e^{-i\lambda_{LH}\gamma^{-1}_c(t-t_z)},$$
$$E(t)= \frac{\epsilon}{\sqrt{2}}(\frac{\alpha}{1+\epsilon}
+\frac{\beta}{1-\epsilon}) e^{-i\lambda_{SL}\gamma^{-1}_c(t-t_z)}+
\frac{1}{\sqrt{2}}(\frac{\alpha}{1+\epsilon}
-\frac{\beta}{1-\epsilon}) e^{-i\lambda_{LH}\gamma^{-1}_c(t-t_z)},$$
$$F(t)=\frac{1}{2}[(\alpha+\beta
\frac{1+\epsilon}{1-\epsilon})e^{-i\lambda_{SL}\gamma^{-1}_c
(t-t_z)}+(\alpha-\beta \frac{1+\epsilon}{1-\epsilon})
e^{-i\lambda_{LH} \gamma^{-1}_c (t-t_z)}],$$
$$G(t)=\frac{1}{2}[(\alpha \frac{1-\epsilon}{1+\epsilon}+\beta
)e^{-i\lambda_{SL} \gamma^{-1}_c (t-t_z)}-(\alpha
\frac{1-\epsilon}{1+\epsilon} -\beta ) e^{-i\lambda_{LH}
\gamma^{-1}_c (t-t_z)}].$$ \end{widetext}

The state of the three particles is thus
\begin{equation} |\Psi_{cab}(t)\rangle
=|\Psi_c(t)\rangle\otimes|\Psi_{ab}(t)\rangle. \label{cab1}
\end{equation}
We let $a$ and $c$ fly in opposite directions and towards each
other; hence, they collide at a certain position $x$ at a certain
time $t_{x}$. The collision can be represented as a unitary
transformation ${\cal S}$ on $c-a$, in a negligible time duration
$\delta$  much shorter than the lifetimes of weak decay. After the
$c-a$ collision, the state of the three kaons can be written as
\begin{equation}
\begin{array}{lll} |\Psi_{cab}(t_x+\delta)\rangle & = &
\frac{M(t_x)}{2}\{\sqrt{2}F(t_x) {\cal
S}|\phi_1\rangle_{ca}|\bar{M}^0\rangle_b \\
&&-\sqrt{2}G(t_x) {\cal
S}|\phi_2\rangle_{ca}|M^0\rangle_b \\
&& - {\cal
S}|\phi_3\rangle_{ca}[F(t_x)|M^0\rangle_b-G(t_x)|\bar{M}^0\rangle_b] \\
&& - {\cal
S}|\phi_4\rangle_{ca}[F(t_x)|M^0\rangle_b+G(t_x)|\bar{M}^0\rangle_b]\},
\end{array}
\label{en22}
\end{equation}
where the $|\phi_i\rangle$ are eigenstates of parity $P$,
strangeness $S$ and isospin $I$: $|\phi_1\rangle \equiv
|M^0M^0\rangle$ with $P=1$, $S=2$ and  $I=1$; $|\phi_2\rangle
\equiv|\bar{M}^0\bar{M}^0\rangle$ with $P=1$, $S=-2$ and $I=1$;
$|\phi_3\rangle \equiv |\Psi_+\rangle \equiv
\frac{1}{\sqrt{2}}(|M^0\rangle|\bar{M}^0\rangle +
|\bar{M}^0\rangle|M^0\rangle)$,  with $P=1$, $S=0$ and $I=1$.
Furthermore, $|\phi_4\rangle \equiv|\Psi_-\rangle$ with $P=-1$,
$S=0$ and $I=0$. ${\cal S}|\phi_i\rangle$ is also an eigenstate of
$S$, $P$ and $I$ with the same eigenvalue as $|\phi_i\rangle$,
because ${\cal S}$ conserves $S$, $P$ and $I$, as governed by the
strong interaction.

The outcomes of the $c-a$ collision are detected by using
interaction with nuclear matter. Hence the state is projected in the
basis $\{{\cal S}|\phi_i\rangle\}$. Conditioned on this projection,
$b$ is known to be, correspondingly, in one of the four states
$|\bar{M}^0\rangle_b$, $|M^0\rangle_b$,
$[F(t_x)|M^0\rangle_b-G(t_x)|\bar{M}^0\rangle_b]/\sqrt{|F(t_x)|^2+|G(t_x)|^2}$,
$[F(t_x)|M^0\rangle_b+G(t_x)|\bar{M}^0\rangle_b]/\sqrt{|F(t_x)|^2+|G(t_x)|^2}$.
If and only if the outgoing particles of the $c-a$ collision are
detected to be with $P=-1$, $S=0$ and $I=0$, then is the $b$
particle retained and known to be in state
$[F(t_x)|M^0\rangle_b+G(t_x)|\bar{M}^0\rangle_b]/\sqrt{|F(t_x)|^2+|G(t_x)|^2}$,
which is just the state of $c$ before collision. This completes the
teleportation from $c$ to $b$. The four possible states of $b$ at
$t_x+\delta$, after knowing the four possible outcomes of the $c-a$
collision, can be verified by measuring the flavor ratio of $b$, as
suggested in Ref.~\cite{shi}.

Now we review entanglement swapping. In addition to
$|\Psi_-\rangle_{ab}$ generated at time $t=0$, another meson pair
$d$ and $c$ is generated as $|\Psi_-\rangle_{dc}$ at $t_z$.
Consequently,
\begin{equation} |\Psi_{dcab}(t)\rangle =
M'(t-t_z)M(t)|\Psi_-\rangle_{dc}|\Psi_-\rangle_{ab},\label{prod}
\end{equation}
where $M'(t-t_z) = \exp
[-i(\lambda_{SL}+\lambda_{LH})\gamma^{-1}_{d}(t-t_z)]$, supposing
$\gamma_c=\gamma_d$. Let $c$ and $a$ fly towards each other to
collide at a certain time $t_x$. Within a negligible time interval
$\delta$, the collision brings about a unitary transformation ${\cal
S}$ on $c-a$; therefore \begin{equation} \begin{array}{lll}
|\Psi_{dcab}(t_x+\delta)\rangle &
=&\frac{M'(t_x-t_z)M(t_x)}{2}({\cal
S}|\Psi_+\rangle_{ca}|\Psi_+\rangle_{db} \\&& - {\cal
S}|\Psi_-\rangle_{ca}|\Psi_-\rangle_{db} \\ && -{\cal
S}|M^0M^0\rangle_{ca}|\bar{M}^0\bar{M}^0\rangle_{db}\\&&- {\cal
S}|\bar{M}^0\bar{M}^0\rangle_{ca}|M^0M^0\rangle_{db}),\end{array}
\label{dcab}\end{equation}  where \begin{eqnarray}
|\Psi_+\rangle_{ca} & \equiv &
\frac{1}{\sqrt{2}}(|M^0\rangle_c|\bar{M}^0\rangle_a +
|\bar{M}^0\rangle_c|M^0\rangle_a)\\
&=& \frac{r}{\sqrt{2}}(|M_{SL}\rangle_c|M_{SL}\rangle_a-
|M_{LH}\rangle_c|M_{LH}\rangle_a).
\end{eqnarray}
Then, in measuring parity $P$, strangeness $S$ and isospin $I$ of
the outgoing particles from the $c-a$ collision, $c$ and $a$ are
projected to one of the four states ${\cal S}|\Psi_{+}\rangle_{ca}$,
${\cal S}|\Psi_-\rangle_{ca}$, ${\cal S}|M^0M^0\rangle_{ca}$ and
${\cal S}|\bar{M}^0\bar{M}^0\rangle_{ca}$. Correspondingly, $d$ and
$b$ are projected to $|\Psi_{+}\rangle_{db}$, $|\Psi_-\rangle_{db}$,
$ |\bar{M}^0\bar{M}^0\rangle_{db}$  and $|M^0M^0\rangle_{db}$,
respectively. Accordingly one chooses to retain or abandon the $b$
particle. The success of entanglement swapping can be verified by
measuring the flavor asymmetry between the $d$ and $b$ particles, as
suggested in Ref.~\cite{shi}.

\section*{4 CP measurement}

\subsection{4.1 Teleportation}

Now we propose that the effect of teleportation and entanglement
swapping can both be verified by measurement in the CP basis. First
we discuss the verification of the effect of teleportation. In the
teleportation scheme outlined above,  the state of $b$, after
knowing the outcomes of the $c-a$ collision, can be verified in the
CP basis. One measures the ratio $\eta$ between the probabilities
for $b$ to be in $|M_{+}\rangle$ and in $|M_{-}\rangle$. For
$$|\Psi(t \geq t_x+\delta)\rangle_b = u_+(t) |M_{+}\rangle_b +
u_-(t)|M_{-}\rangle_b,$$ $$\eta(t)\equiv
|\frac{u_+(t)}{u_-(t)}|^2.$$ Many runs of the procedure, or many
copies of $b$ particles in a beam, are needed to determine this
quantity.

If irrespective of the outcome of the $c-a$ collision,  $b$
particles in different runs of the experiment are all considered in
measuring $\eta(t)$, then $\eta(t)$ should be calculated by using
$|\Psi_{cab}(t)\rangle$, given in Eq.~(\ref{cab1}). Because $b$ is
maximally entangled with $a$, it can be found that $\eta_b(t)=1$. In
contrast, if only $b$ particles in those runs of the experiment with
a certain projection result of $c-a$ are considered in measuring
$\eta(t)$, then $\eta(t)$ is calculated by using the corresponding
projected state of $b$, as seen in Eq.~(\ref{en22}). Denote the
state of $b$ following the projection as
$\alpha'|M^0\rangle+\beta'|\bar{M}^0\rangle$. Its subsequent
evolution in the CP basis is then similar to Eq.~(\ref{cp}), with
$t_z$ substituted for by $t_x+\delta$, $\gamma_c$ by $\gamma_b$,
$\alpha$ by $\alpha'$ and $\beta$ by $\beta'$. It can be found that
$$\eta_b(t)= |\frac{(\frac{\alpha'}{1+\epsilon} +\frac{\beta'}{1-\epsilon})
+ \epsilon(\frac{\alpha'}{1+\epsilon} -\frac{\beta'}{1-\epsilon})
e^{-i\Delta\lambda\tau}} { \epsilon(\frac{\alpha'}{1+\epsilon}
+\frac{\beta'}{1-\epsilon}) + (\frac{\alpha'}{1+\epsilon}
-\frac{\beta'}{1-\epsilon}) e^{-i\Delta\lambda\tau}}|^2,$$ where
$\tau=\gamma^{-1}_b(t-t_x-\delta)$ and
$\Delta\lambda=\lambda_{LH}-\lambda_{SL}=\Delta m-i\Delta\Gamma$,
with $\Delta m = m_{LH} - m_{SL}$ and $\Delta \Gamma =
\Gamma_{LH}-\Gamma_{SL}/2$.

For each of the four projection cases, $\eta_b(t)$ is very different
from $1$. If $c-a$ projects to ${\cal S}|M^0M^0\rangle$, then
$$\eta_b= |\frac{1-\epsilon
e^{-i\Delta\lambda\tau}}{\epsilon-e^{-i\Delta\lambda\tau}}|^2.$$ If
$c-a$ projects to ${\cal S}|\bar{M}^0\bar{M}^0\rangle$, then
$$\eta_b= |\frac{1+\epsilon
e^{-i\Delta\lambda\tau}}{\epsilon+e^{-i\Delta\lambda\tau}}|^2. $$ If
$c-a$ projects to ${\cal S}|\Psi_+\rangle$, then $$\eta_b(t)=
|\frac{(\frac{F(t_x)}{1+\epsilon} -\frac{G(t_x)}{1-\epsilon}) +
\epsilon(\frac{F(t_x)}{1+\epsilon} +\frac{G(t_x)}{1-\epsilon})
e^{-i\Delta\lambda\tau}} { \epsilon(\frac{F(t_x)}{1+\epsilon}
-\frac{G(t_x)}{1-\epsilon}) + (\frac{F(t_x)}{1+\epsilon}
+\frac{G(t_x)}{1-\epsilon}) e^{-i\Delta\lambda\tau}}|^2.$$ If $c-a$
projects to ${\cal S}|\Psi_-\rangle$, i.e. the teleportation is
successful, then $$\eta_b(t)= |\frac{(\frac{F(t_x)}{1+\epsilon}
+\frac{G(t_x)}{1-\epsilon}) + \epsilon(\frac{F(t_x)}{1+\epsilon}
-\frac{G(t_x)}{1-\epsilon}) e^{-i\Delta\lambda\tau}} {
\epsilon(\frac{F(t_x)}{1+\epsilon} +\frac{G(t_x)}{1-\epsilon}) +
(\frac{F(t_x)}{1+\epsilon} -\frac{G(t_x)}{1-\epsilon})
e^{-i\Delta\lambda\tau}}|^2.$$

\subsection*{4.2 Entanglement swapping}

Now we turn to entanglement swapping, which  can also be verified by
measurement in the CP basis. One can measure CP asymmetry between
$b$ and $d$, defined as
$$A_{cp}(t)=\frac{p_{d,cp}(t)-p_{s,cp}(t)}{p_{d,cp}(t)+p_{s,cp}(t)},$$
where $p_{d,cp}(t)$ and $p_{s,cp}(t)$ are, respectively, the
probabilities for $b$ and $d$ to have different and the same values
of CP~\cite{cern}. Many runs of the experiment, or many copies of
the particles in a beam, are needed to experimentally determine
$A_{cp}(t)$. If all the $d-b$ pairs in different runs are
considered, irrespective of the projection results of $c-a$, it can
be found that $A_{cp}(t)=0$, as calculated from
$|\Psi_{dcab}(t)\rangle$, by (\ref{r}), (\ref{cp}) and (\ref{prod}).
In contrast, if only those $d-b$ pairs corresponding to a certain
projection result of the $c-a$ collision are considered, then
$A_{cp}(t)$ is calculated by using the corresponding projected state
of $d$ and $b$, as determined from $|\Psi_{dcab}(t_x+\delta)\rangle$
given in Eq.~(\ref{dcab}). In the following, we give calculations
for the four cases of projection.

{\em Case 1}. If at $t=t_x+\delta$, $c$ and $a$ are projected to
${\cal S}|\Psi_+\rangle_{ca}$, then
\begin{widetext}
\begin{eqnarray}
|\Psi(t\geq t_x+\delta)\rangle_{db}  =
\frac{r}{\sqrt{2}}[e^{-i(\lambda_{SL}\tau_d+\lambda_{SL}\tau_b)}|M_{SL}\rangle_d|M_{SL}\rangle_b
- e^{-i(\lambda_{LH}\tau_d+\lambda_{LH}\tau_b)}
|M_{LH}\rangle_d|M_{LH}\rangle_b ]  \nonumber\\
= \frac{1}{\sqrt{2}(1-\epsilon^2)}[\epsilon(f_1-f_2)
(|M_-M_+\rangle+|M_+M_-\rangle) +
(\epsilon^2f_1-f_2)|M_-M_-\rangle+(f_1-\epsilon^2
f_2)|M_+M_+\rangle], \nonumber
\end{eqnarray}
where  $\tau_d=\gamma^{-1}_d(t-t_x-\delta)$,
$\tau_b=\gamma^{-1}_b(t-t_x-\delta)$,
$f_1=e^{-i(\lambda_{SL}\tau_d+\lambda_{SL}\tau_b)}$,
$f_2=e^{-i(\lambda_{LH}\tau_d+\lambda_{LH}\tau_b)}$. As $d$ and $b$
originate from different sources, $\gamma_d$ and $\gamma_b$ may be
different. Consequently
$$A_{cp}(t \geq t_x+\delta)=
\frac{-(1+a_{\epsilon0})^2(1+e^{-\Delta \Gamma(\tau_d+\tau_b)})-
2a_{\epsilon +}^2e^{-\frac{\Delta\Gamma}{2}(\tau_d+\tau_b)}
\cos\Delta m(\tau_d+\tau_b)}{1+e^{-\Delta \Gamma(\tau_d+\tau_b)}-2
a_{\epsilon}^2e^{-\frac{\Delta\Gamma}{2}(\tau_d+\tau_b)} \cos\Delta
m(\tau_d+\tau_b)} , \nonumber$$ where we have used rephase-invariant
CP-violating observables~\cite{WP}
\begin{eqnarray}
a_{\epsilon} = \frac{2Re\epsilon}{1 + |\epsilon|^2}, \quad
a_{\epsilon+} = \frac{2Im\epsilon}{1 + |\epsilon|^2}, \nonumber
\end{eqnarray}
which characterize indirect CP violation  and mixing-induced CP
violation, respectively, and the third quantity $ a_{\epsilon0} = -
2|\epsilon|^2/(1 + |\epsilon|^2)$, which is related to
$a_{\epsilon}$ and  $a_{\epsilon+}$ through $(1+a_{\epsilon0})^2 +
a_{\epsilon+}^2 = 1- a_{\epsilon}^2$~\cite{WP}.  As CP violation is
small in the neutral meson system, $a_{\epsilon} << 1$,
$a_{\epsilon+}<< 1$, and thus $a_{\epsilon0}<<1$. Therefore $A_{cp}$
is close to $-1$.

{\em Case 2}. If at $t=t_x+\delta$, $c$ and $a$ are projected to
${\cal S}|\Psi_-\rangle_{ca}$, i.e. the entanglement swapping is
successful, then for $t \geq t_x+\delta$,
\begin{eqnarray}
|\Psi(t\geq t_x+\delta)\rangle_{db}   = \frac{r}{\sqrt{2}}
[e^{-i(\lambda_{LH}\tau_d+\lambda_{SL}\tau_b)}|M_{LH}\rangle_d|M_{SL}\rangle_b
- e^{-i(\lambda_{SL}\tau_d+\lambda_{LH}\tau_b)}
|M_{SL}\rangle_d|M_{LH}\rangle_b ],  \nonumber\\
= \frac{1}{\sqrt{2}(1-\epsilon^2)}[(g_1-\epsilon^2g_2)|M_-M_+\rangle
+ \epsilon(g_1-g_2)(|M_-M_-\rangle+|M_+M_+\rangle)+(\epsilon^2
g_1-g_2)|M_+M_-\rangle], \nonumber
\end{eqnarray}   where
$g_1=e^{-i(\lambda_{LH}\tau_d+\lambda_{SL}\tau_b)}$,
$g_2=e^{-i(\lambda_{SL}\tau_d+\lambda_{LH}\tau_b)}$.  Consequently
$$A_{cp}(t \geq t_x+\delta)=
\frac{(1 + a_{\epsilon0} )^2(1+e^{-\Delta \Gamma(\tau_d-\tau_b)})+
2a_{\epsilon+}^2e^{-\frac{\Delta\Gamma}{2}(\tau_d-\tau_b)}
\cos\Delta m(\tau_d-\tau_b)}{1+e^{-\Delta
\Gamma(\tau_d-\tau_b)}-2a_{\epsilon}^2e^{-\frac{\Delta\Gamma}{2}(\tau_d-\tau_b)}
\cos\Delta m(\tau_d-\tau_b)} , \nonumber$$ which is close to $1$.

{\em Case 3}. If at $t=t_x+\delta$, $c$ and $a$ are projected to
${\cal S}|M^0M^0\rangle_{ca}$, then for $t \geq t_x+\delta$,
\begin{eqnarray}
|\Psi(t\geq t_x+\delta)\rangle_{db}&=&
\frac{1+|\epsilon|^2}{2(1-\epsilon)^2}
[f_1|M_{SL}\rangle_d|M_{SL}\rangle_b
 - g_2 |M_{SL}\rangle_d|M_{LH}\rangle_b
 -g_1|M_{LH}\rangle_d|M_{SL}\rangle_b +f_2
|M_{LH}\rangle_d|M_{LH}\rangle_b ]  \nonumber\\&
=&\frac{1}{2(1-\epsilon)^2}[(f_1-\epsilon g_2-\epsilon g_1
+\epsilon^2 f_2)|M_+M_+\rangle+ (\epsilon f_1 -g_2-\epsilon^2 g_1
+\epsilon f_2) |M_+M_-\rangle \nonumber \\ &&+(\epsilon f_1
-\epsilon^2 g_2- g_1 +\epsilon f_2) |M_-M_+ \rangle+ (\epsilon^2
f_1-\epsilon g_2-\epsilon g_1 +f_2)|M_-M_-\rangle]. \nonumber
\end{eqnarray} Consequently,
$A_{cp}(t \geq t_x+\delta)= \{2a_{\epsilon+}[(1-|\epsilon|^2)
e^{-2\Gamma_{SL}\tau_d}-e^{-2\Gamma_{LH}\tau_d}]
e^{-(\Gamma_{LH}+\Gamma_{SL})\tau_b}\sin\Delta
m\tau_b+2a_{\epsilon+}[
e^{-2\Gamma_{SL}\tau_b}-(1-|\epsilon|^2)e^{-2\Gamma_{LH}\tau_b}]
e^{-(\Gamma_{LH}+\Gamma_{SL})\tau_d}\sin\Delta
m\tau_d+2a_{\epsilon}|\epsilon|^2(e^{-[2\Gamma_{SL}\tau_b+
(\Gamma_{LH}+\Gamma_{SL})\tau_d]}\cos\Delta m\tau_d
+e^{-[2\Gamma_{LH}\tau_d+
(\Gamma_{LH}+\Gamma_{SL})\tau_b]}\cos\Delta
m\tau_b)+4a_{\epsilon+}^2(1+|\epsilon|^2)e^{-(\Gamma_{LH}+\Gamma_{SL})(\tau_d+\tau_b)}
\cos\Delta m\tau_d\cos\Delta m\tau_b\}/\{
(1+|\epsilon|^2)(e^{-\Gamma_{SL}(\tau_d+\tau_b)}+
e^{-(\Gamma_{SL}\tau_d+\Gamma_{LH}\tau_b)}+e^{-\Gamma_{LH}(\tau_d+\tau_b)}
e^{-(\Gamma_{LH}\tau_d+\Gamma_{SL}\tau_b)})-2a_{\epsilon}
[(1+|\epsilon|^2)e^{-2\Gamma_{SL}\tau_d}+e^{-2\Gamma_{LH}\tau_d}]
e^{-(\Gamma_{LH}+\Gamma_{SL})\tau_b}\cos\Delta m\tau_b
-2a_{\epsilon}
[(1+|\epsilon|^2)e^{-2\Gamma_{LH}\tau_b}+e^{-2\Gamma_{SL}\tau_b}]
e^{-(\Gamma_{LH}+\Gamma_{SL})\tau_d}\cos\Delta m\tau_d +
2a_{\epsilon+}|\epsilon|^2
(e^{-[2\Gamma_{SL}\tau_b+(\Gamma_{LH}+\Gamma_{SL})\tau_d]}
\sin\Delta m\tau_d
-e^{-[2\Gamma_{LH}\tau_d+(\Gamma_{LH}+\Gamma_{SL})\tau_b]}
\sin\Delta m\tau_b) +4a_{\epsilon}^2(1+|\epsilon|^2)
e^{-(\Gamma_{LH}+\Gamma_{SL})(\tau_d+\tau_b)}\cos\Delta
m\tau_d\cos\Delta m\tau_b\}, $ which has been written in terms of
rephase-invariant quantities $a_{\epsilon}$ and $a_{\epsilon+}$, as
well as $|\epsilon|^2$, which can be substituted as $-a_{\epsilon
0}/(2+a_{\epsilon 0})$. In this case, $A_{cp}$ is close to $0$.

{\em Case 4}. If at $t=t_x+\delta$, $c$ and $a$ are projected to
${\cal S}|\bar{M}^0\bar{M}^0\rangle_{ca}$, then for $t \geq
t_x+\delta$,
\begin{eqnarray} |\Psi(t\geq t_x+\delta)\rangle_{db}&=&
\frac{1+|\epsilon|^2}{2(1+\epsilon)^2}
[f_1|M_{SL}\rangle_d|M_{SL}\rangle_b
 +g_2 |M_{SL}\rangle_d|M_{LH}\rangle_b
 +g_1|M_{LH}\rangle_d|M_{SL}\rangle_b +f_2
|M_{LH}\rangle_d|M_{LH}\rangle_b ]  \nonumber\\&
=&\frac{1}{2(1+\epsilon)^2}[(f_1+\epsilon g_2+\epsilon g_1
+\epsilon^2 f_2)|M_+M_+\rangle+ (\epsilon f_1 +g_2+\epsilon^2 g_1
+\epsilon f_2) |M_+M_-\rangle \nonumber\\ &&+(\epsilon f_1
+\epsilon^2 g_2+ g_1 +\epsilon f_2) |M_-M_+ \rangle+ (\epsilon^2
f_1+\epsilon g_2+\epsilon g_1 +f_2)|M_-M_-\rangle]. \nonumber
\end{eqnarray}
Consequently,  $A_{cp}(t \geq t_x+\delta)=
\{2a_{\epsilon+}(1-|\epsilon|^2)[
(e^{-2\Gamma_{LH}\tau_b}-e^{-2\Gamma_{SL}\tau_b})
e^{-(\Gamma_{LH}+\Gamma_{SL})\tau_d}\sin\Delta m\tau_d+
(e^{-2\Gamma_{LH}\tau_d}-e^{-2\Gamma_{SL}\tau_d})
e^{-(\Gamma_{LH}+\Gamma_{SL})\tau_b}\sin\Delta
m\tau_b]+4a_{\epsilon+}^2(1+|\epsilon|^2)
e^{-(\Gamma_{LH}+\Gamma_{SL})(\tau_d+\tau_b)} \cos\Delta
m\tau_d\cos\Delta m\tau_b\}/\{
(1+|\epsilon|^2)(e^{-\Gamma_{SL}(\tau_d+\tau_b)}+
e^{-(\Gamma_{SL}\tau_d+\Gamma_{LH}\tau_b)}+e^{-\Gamma_{LH}(\tau_d+\tau_b)}
e^{-(\Gamma_{LH}\tau_d+\Gamma_{SL}\tau_b)})+2a_{\epsilon}
(1+|\epsilon|^2)[(e^{-2\Gamma_{SL}\tau_d}+e^{-2\Gamma_{LH}\tau_d})
e^{-(\Gamma_{LH}+\Gamma_{SL})\tau_b}\cos\Delta m\tau_b
+(e^{-2\Gamma_{LH}\tau_b}+e^{-2\Gamma_{SL}\tau_b})
e^{-(\Gamma_{LH}+\Gamma_{SL})\tau_d}\cos\Delta m\tau_d +
4a_{\epsilon}^2(1+|\epsilon|^2)e^{-(\Gamma_{LH}+\Gamma_{SL})(\tau_d+\tau_b)}
\cos\Delta m\tau_d\cos\Delta m\tau_b\}, $ which has been written in
terms of the rephase-invariant quantities $a_{\epsilon}$ and
$a_{\epsilon+}$, as well as $|\epsilon|^2$, which can be substituted
for by $-a_{\epsilon 0}/(2+a_{\epsilon 0})$. In this case, $A_{cp}$
is close to $0$.
\end{widetext}

Therefore, no matter whether the two lifetimes are considerably
different from each other, and no matter whether CP is violated, the
success of the entanglement swapping can be clearly distinguished
from the other three cases of projection, as well as from the case
of no projection, by measuring the CP asymmetry. Obviously, CP
asymmetry can also be similarly used in verifying the general
teleportation, in which the teleported kaons are entangled
arbitrarily with other particles in an unknown way~\cite{shi}.

Certainly, CP asymmetry can also be used for verifying the
Einstein-Podolsky-Rosen state $|\Psi_-\rangle$ generated in $e^+e^-$
or $p\bar{p}$ collisions, as in the CPLEAR and BELLE, where flavor
asymmetries were measured. However, it should be noted that these
asymmetry quantities, including the those measured in CPLEAR and
BELLE experiments, are by no means rigorous proofs of entanglement,
because it is easy to construct separated states with the same
asymmetry as an entangled state.

\section*{5 Summary}

To summarize, we have described a  scheme of using measurement in
the CP basis to verify  quantum teleportation or entanglement
swapping in terms of neutral pseudoscalar mesons. This method has
several advantages. It is rigorous, and it remains valid in the
presence of CP violation. It also works efficiently, clearly
distinguishing the success case in teleportation or entanglement
swapping from other cases. Furthermore, this method is much simpler
than the flavor measurement using the  strong basis. The latter
needs nuclear matter to interact with the particles to be detected,
while in the present method, only the decay modes need to be
determined. This aspect brings high energy quantum information
manipulation in general, and quantum teleportation or entanglement
swapping in particular, closer to actual experimental
implementation, which, however, should still need to overcome other
serious difficulties, e.g. precise control of the timing, careful
determination of collision outcomes, etc. Finally we emphasize that
CP measurements can also be used in general tests of quantum
mechanical effects in particle physics.

{\em Acknowledgements.} This work is supported by National Science
Foundation of China (Grant No. 10674030), Shanghai Pujiang Project
(Grant No. 06PJ14013) and Shanghai Shuguang Project (Grant No.
07S402).

\end{document}